# A Generalized Accelerated Failure Time Model to Predict Restoration Time from Power Outages

**Tasnuba Binte Jamal (Corresponding author)**

Ph.D. Student

Department of Civil, Environmental, and Construction Engineering

University of Central Florida

12800 Pegasus Drive, Orlando, FL 32816, USA

Email: tasnubajamal@Knights.ucf.edu

**Samiul Hasan, Ph.D.**

Associate Professor

Department of Civil, Environmental, and Construction Engineering

University of Central Florida

12800 Pegasus Drive, Orlando, FL 32816, USA

Email: Samiul.Hasan@ucf.edu

**ACKNOWLEDGMENTS**

The authors are grateful to the U.S. National Science Foundation for the grant CMMI-1832578 to support the research presented in this article. However, the authors are solely responsible for the findings presented here.

**ABSTRACT**

Major disasters such as wildfire, tornado, hurricane, tropical storm, flooding cause disruptions in infrastructure systems such as power outage, disruption to water supply system, wastewater management, telecommunication failures, and transportation facilities. Disruptions in electricity infrastructures has a negative impact on every sector of a region, such as education, medical services, financial, recreation, and so on. In this study, we introduce a novel approach to investigate the factors which can be associated with longer restoration time of power service after a hurricane. We consider three types of factors (hazard characteristics, built-environment characteristics, and socio-demographic factors) such as wind speed, urban classification, ethnicity that might be associated with longer restoration times of power outages during a hurricane. Considering restoration time as the dependent variable and utilizing a comprehensive set of county-level data, we have estimated a Generalized Accelerated Failure Time (GAFT) that accounts for spatial dependence among observations for time to event data. Considering spatial correlation has improved the model fit by 12% from a model without considering spatial dependence in time to event data. Using GAFT model and Hurricane Irma's impact on Florida as a case study, we examined: (1) differences in electric power outages and restoration rates among different types of power companies: investor-owned power companies, rural and municipal cooperatives; (2) the relationship between the duration of power outage and power system variables; and (3) the relationship between the duration of power outage and socioeconomic attributes. We have found that factors such as maximum sustained wind speed, percentage of customers facing power outage, percentage of customers served by investor-owned power company, median household income and number of power plants are strongly associated with restoration time. The findings of this study indicate that counties with a higher percentage of customers served by investor-owned electric companies and lower median household income, faced power outage for a longer time. This paper identifies the key factors in predicting the restoration time of hurricane-induced power outages. This study will help estimate the restoration time after a hurricane allowing emergency management professionals to properly adopt strategies required for restoration process.

# 1. INTRODUCTION

Hurricanes have become more frequent and intense due to climate change and global warming issues. Hurricane induced damages have significantly increased because of major landfalls in recent years (Grenier et al., 2020). For instance, Hurricane Irma caused a damage of about $50 billion (Cangialosi et al., 2017) and significant disruptions in infrastructure systems. After Hurricane Irma, more than 6.2 million customers lost power including 850,000 customers from Orange, Seminole, Lake, and Osceola counties in Florida (Gillespie et al., 2017). Similarly, 8.5 million customers lost power during Hurricane Sandy (Alemazkoor et al., 2020). Sustained winds and excessive amount of precipitation/flooding during hurricanes cause disruptions to infrastructure systems such as power outage, disruptions in water supply and wastewater systems, telecommunication failures, and transportation systems disruptions. Local communities depend on these systems to a great extent and failures in such infrastructure systems highly affect their daily activities.

Moreover, infrastructure systems have become highly interconnected and interdependent (Grafius et al., 2020; Rinaldi et al., 2001). After Hurricane Sandy, damages in electricity stations significantly affected the functions of transportation facilities (Haraguchi & Kim, 2016). Power outages hampered the restoration of subway services in the New York City as trains could not run without power restoration. Due to the interconnected and interdependent relationships among infrastructure systems, the restoration process of a system is further delayed when other systems are disrupted. As a result, infrastructure services become unavailable and the quality of life of the served population decreases. Among all types of infrastructure disruptions, a disruption in the electricity power infrastructure is the most significant one. Power outages have significant negative impacts on every sector of a region such as business, agriculture, and economic growth (Koks et al., 2019; Kuntke et al., 2022; Mukherje & Hastak, 2018).

To enhance community resilience against an extreme event, faster restoration from power outages is necessary for recovery efforts. For this reason, six steps are proposed for power restoration process including restoration at power plants, at transmission lines, at substations, for essential services, in large service areas, and at individual home (Edison Electric Institute, 2019). To date, most of the works in infrastructure disruption network analysis and modeling frameworks have involved the first three steps (Ouyang & Wang, 2015). However, a holistic approach for the restoration process at the last two steps (restoration in large service areas and at individual home) is essential to understand power disruption patterns and durations at regional and household levels.

Previous studies investigated the durations of power outages after hurricanes at a regional level. (Liu et al., 2007) applied Accelerated Failure Time (AFT) model to understand restoration time of power outage. While this approach provides useful insights for time to event data analysis, it ignores spatial clustering of restoration time for power outage. To mitigate this problem, (Mitsova et al., 2018) applied

Spatial Autoregressive (SAR) model to understand restoration time of power outage at a county level. While this model considers the spatial autocorrelation among observations, it does not provide useful insights for time to event data analysis. The overall restoration time of power outages can be explained by a set of easily available independent variables using a Generalized Accelerated Failure Time (GAFT) model that allows both time to event data analysis and spatial autocorrelation (non-independence of the observations) (Zhou et al., 2020).

In this study, we investigate the spatial extent and correlation of restoration time of electricity disruptions across different counties after a hurricane by applying a spatial clustering approach. We also develop a statistical model (GAFT model) considering a range of variables including hazard, built environment, and socio-demographic characteristics to identify the factors associated with longer restoration time for power outage at a county level after a hurricane. If restoration time can be reliably predicted, households may plan for alternatives of existing power services during disruptions. At the same time, when policymakers and stakeholders better understand the factors associated with longer disruptions, they can allocate resources to manage restoration processes, reduce restoration time, and mitigate the negative impacts of longer restoration times from power outages.

Our study has the following contributions:

- We investigate the spatial distribution of restoration time of electricity disruption of a region during a hurricane using a statistical clustering approach.
- We develop Generalized Accelerated Failure Time (GAFT), a statistical model to investigate the association between restoration time of power outage and a wide range of variables including hazard, built-environment factors, and socio-demographic characteristics of the regions accounting for spatial dependence of the data. While power outage has been studied from the perspective of time to event data analysis and considering the spatial dependence of the observations separately, we add a new dimension by developing Generalized Accelerated Failure Time (GAFT) model which can account both for time to event data and spatial dependence of the data.

**2. LITERATURE REVIEW**

Infrastructure systems are highly interconnected and interdependent and the declined performance in one system significantly affects other systems (Hasan & Foliente, 2015; Grafius et al., 2020; Rinaldi et al., 2001). In recent times, there has been an increased interest in studying the impact of power outage on the performance of other infrastructure systems after extreme events and how to enhance the resilience of such interdependent systems.

For instance, previous studies focused on power-water network disruptions for natural hazards and suggested possible solutions to ensure a resilient power-water distribution system after a hurricane (Almoghathawi et al., 2019; Kong et al., 2021; Najafi et al., 2019, 2020). Disruptions in electricity and petroleum infrastructures had negative impacts on health care services, public transportation systems in the New York metropolitan area after Hurricane Sandy (Haraguchi & Kim, 2016). Traffic congestion increased three to four times due to the power outages after hurricane Isaac (Miles & Jagielo, 2014). Previous studies developed models to assess the resilience of interdependent traffic- power systems and to determine the parameters to quantify the resilience of transportation systems against hurricanes and other natural disasters (Kocatepe et al., 2018; Zou & Chen, 2020) and (Ahmed & Dey, 2020). Zou & Chen (2020) proposed strategies to improve the resilience of traffic-power systems against a hurricane. Ouyang & Wang (2015) modeled for the resilience optimization of interdependent infrastructures. The consequences of the interdependencies in infrastructure failures starting from a given outage were analyzed by considering severity, duration, spatial extent, and the number of people affected by a disruption (Mcdaniels et al., 2007). Kong et al. (2021) calculated the infrastructure efficiency by removing different percentage of nodes in the system for both power and water systems. Previous studies also investigated the societal, mental, and economic impact of power disruption (Dargin & Mostafavi, 2020; Stock et al., 2021) along with interdependency analysis among the infrastructure systems. Studies have explored recovery strategies and efficiency (Ge et al., 2019; Loggins et al., 2019) as well.

Most of the above studies considered infrastructure disruptions at an infrastructure facility level, analyzing how a disruption in an electric power plant affects a water treatment plan or water distribution systems or a gas station after a disaster. These studies focused on the restoration at power plants, transmission lines, and substations. Also, these studies focused on the impact of power outage. For example, previous research mainly focused on how and to what extent other facilities are disrupted when an extreme event takes place causing power outages in a region. Few studies investigated power service disruptions at a local (e.g., county) level and the time required for the restoration process. Restoration time from power outages needs more attention along with impact analysis at a local scale.

Researchers have developed models to identify the contributing factors toward power outage following a disaster. Liu et al. (2007) developed an Accelerated Failure Time (AFT) model for determining the time required for the restoration of power outage after an extreme hazard, considering hurricane and snowstorm. Nateghi et al., (2011) compared different models such as Accelerated Failure Time model, Cox proportional hazard model, regression trees, Bayesian additive regression trees (BART), and multivariate additive regression splines and found that BART performs the best. Models based on various Geographic Information System databases were developed to determine where outages are most likely to occur by Liu et al. (2005). Han et al. (2009) considered hurricane characteristics, land cover, power system data to analyze the number of outage and spatial distribution of the power outage using negative binomial generalized linear model for the Gulf Coast region of the USA. Quiring et al. (2011) included soil characteristics and suggested that these variables can implicitly inform about the likelihood of trees being uprooted. McRoberts et al., (2018) showed that the inclusion of elevation, land cover, soil, precipitation, and vegetation characteristics improved the accuracy of previously established statistical model by 17%. However, sociodemographic characteristics and social vulnerability of population were not considered in these studies. Dargin & Mostafavi (2020) considered sociodemographic factors of a community and identified which group of people were affected mostly from well-being perspectives due to various infrastructure disruptions after Hurricane Harvey. However, the spatial distribution of the recovery process for a particular disruption like which group of people faced longer disruption was not considered.

Socio-economic and socio-demographic characteristics of the affected regions were considered in previous research. Mitsova et al. (2019) considered characteristics such as age, gender, race, housing tenure, education, and income to identify whether households are recovered or not from power outages after Hurricane Irma. They found that while distributing federal financial assistance, low-income households and minority groups were given less priority. Duffey, (2019) used a wide variety of extreme events, such as hurricanes, wildfires, heavy snowstorms, and devastating cyclones, to calculate recovery times and probabilities of failure to restore. He found that wildfires and hurricanes may have different causes, but the non-restoration probability patterns they produce are identical: a straightforward exponential decline. Lee et al. (2019) studied the disparity in getting social supports (e.g., instrumental, emotional, informational, and outside contact support) considering respondents' sociodemographic characteristics such as education, age, and religion. Results imply that older and less educated people faced constraints in post-disaster support. Previous studies also found that regions served by rural municipalities faced longer disruption for electricity disruption after Hurricane Maria and Irma (Mitsova et al., 2018; O. Román et al., 2019). Using satellite nighttime lights data for Hurricane Maria, O. Román et al. (2019) found that within same urban area, poor residents possess higher risk of power loss and longer disruption time. To determine the relationship between physical and socioeconomic characteristics and the power recovery effort, Azad & Ghandehari, (2021) developed a Quasi-Poisson regression model and found that major challenges to the repair work were poor road infrastructure and economically depressed communities. These studies considered sociodemographic information of the households and regions while analyzing the effects of electricity disruption and time required for recovery operation for power outage. However, these studies did not focus on the distribution of restoration time to get back the power over regions.

In addition, other considerations such as meteorological, housing characteristics etc. are equally important to perceive the dynamics of restoration process for the disaster response community. Watson et al., (2022) developed a machine learning model for predicting the effects of extreme weather events on electrical distribution grids. They found substantial diversity in the meteorological factors that drive the predictions for the most severe events, suggesting that weather hazards are more complex than they are often treated in empirical analyses of their impacts. Wanik et al., (2018) simulated Hurricane Sandy like scenarios in the future to determine the severity of tree-caused outages in Connecticut, with each showing increased winds and higher rain accumulation over the study area as a result of large-scale thermodynamic changes in the atmosphere and track modifications in 2100. Using an ensemble of Weather Research and Forecasting simulations coupled with three machine learning based outage prediction models, they found that future Sandy will lead to a 42%-64% rise in outages. Mukherjee et al. (2018) characterized the key factors of severe weather-induced power outages and found that power outage risk is a function of the type of natural hazard, and investments in operations/maintenance activities (e.g., tree-trimming, replacing old equipment, etc.). These studies found weather impacts on power grids and density of power outages with simulations and machine learning algorithms.

Previous studies used various statistical modeling approaches for estimating power service restoration time. Mitsova et al. (2018) developed a Spatial Autoregressive (SAR) model at a county level for Hurricane Irma to determine the attributes associated with restoration time from power outages. However, like AFT model, SAR model does not provide useful insights for time to event data analysis. We used GAFT model for two reasons: (i) as our dependent variable is restoration time, we considered this to be a time to event data analysis; (ii) as our dependent variable is likely to have a spatial dependency, the GAFT model can be used for modeling both spatial and non-spatial data. Similarly, using a Spatial Autoregressive model, Ulak et al. (2018) included wind speed, infrastructure and transportation characteristics, demographic, and socioeconomic characteristics to predict the number of power outages in the city of Tallahassee of Florida for Hurricane Hermine. Rachunok & Nateghi (2020) considered the spatial distribution of disruptions by demonstrating the network-performance of the power distribution grid's sensitivity to spatial characteristics. However, we should give more emphasis on restoration time rather than outage density. If many customers face power outage after a hurricane but they get back their power services within a short period, it may not hamper much to their business, social, and other daily activities. Besides SAR model, a random forest model is used to predict hurricane-induced power outage durations (Nateghi et al., 2014) and outages (Guikema et al., 2014), which do not provide useful insights for time to event data analysis and spatial clustering of restoration time.

In summary, the following observations can be made. First, two types of dependent variables have been considered in the existing literature: number of outages and duration of outages. Second, duration of outages was examined at different geographical levels ranging from grid sizes to county subdivisions and county levels. Lastly, different types of factors such as meteorological, physical factors, socio-demographic attributes were considered to explain the outage durations developing different machine learning and statistical models. Previously no models have been developed which can account for both 'time' as a dependent variable and spatial autocorrelation. This paper claims a methodological contribution for modeling power outage restoration time by considering spatial dependence and time to event aspects present in the data. This can provide more accurate and reliable predictions of restoration time with significant implications on policymaking related to infrastructure planning and management. More specifically, the following research questions are yet to be answered: (i) *can we implement a statistical model which can account both for time to event data and spatial dependence of the data to predict restoration time from hurricane-induced power outages from a set of common key factors that are publicly available? (ii) can we explain the spatial distribution of restoration time of electricity disruption due to a hurricane using a statistical clustering approach?* As such, the objectives of this study are to understand the spatial clustering patterns of the restoration time of power outages due to a hurricane and to determine the factors associated with prolonged restoration time considering a wide range of variables including hazard, built environment, and socio-demographic characteristics.

## 3. DATA COLLECTION AND PROCESSING

### 3.1 Restoration time

We collected the data for the restoration time of power outages during Hurricane Irma from Florida Today ([data.floridatoday.com/storm-power-outages/](data.floridatoday.com/storm-power-outages/)) for each county of Florida. The plots in Florida Today were drawn using the data from Florida Division of Emergency Management. We used the duration between the time when 20% customers or more of a particular county first lost their electricity services and the time when 20% customers or less were yet to restore their power services (**Fig. 1**). Because it was observed that the counties where less than 20% customers lost power services, did not take long time to get their electricity services back. Such counties are Calhoun and Washington. Besides, in Okaloosa, Santa Rosa, and Walton counties, no customer faced power disruption due to Hurricane Irma. In most of the counties (19 counties), it took 3 days to restore the power service for at least 80% of their customers (**Fig. 2**).

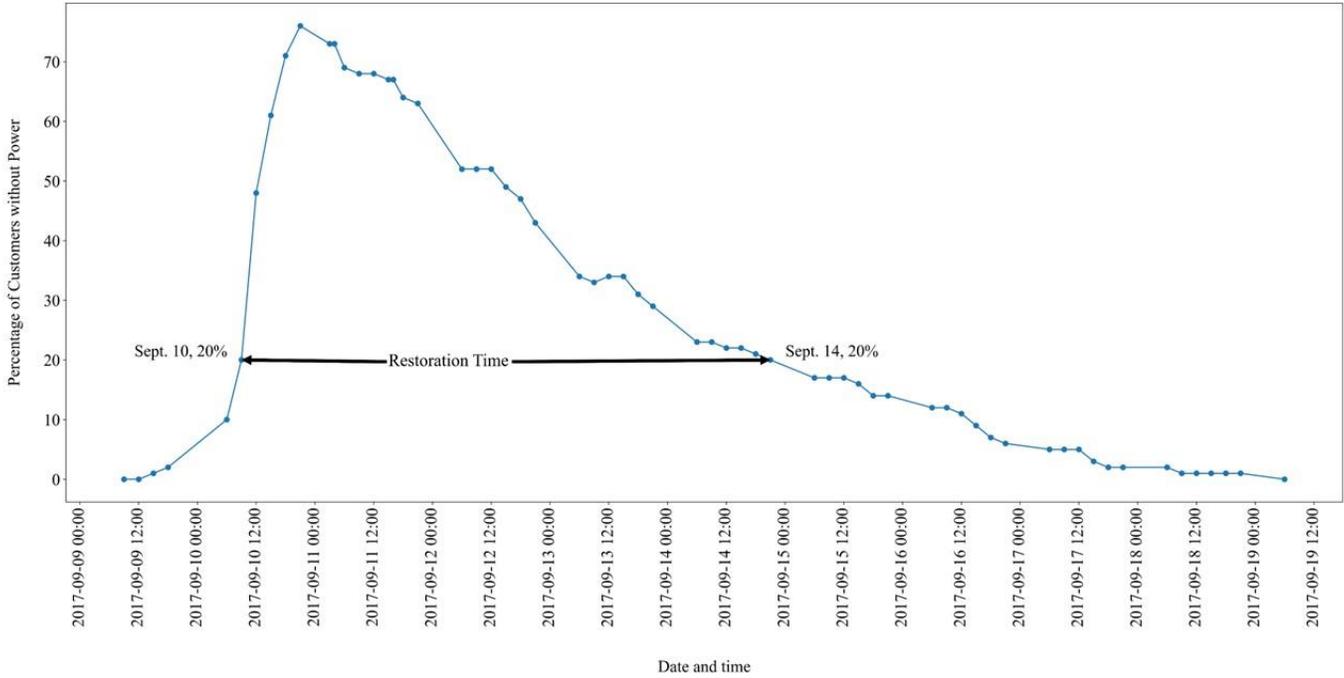

**Fig. 1.** Power outage for Broward County after Hurricane Irma and considered restoration time in this study.

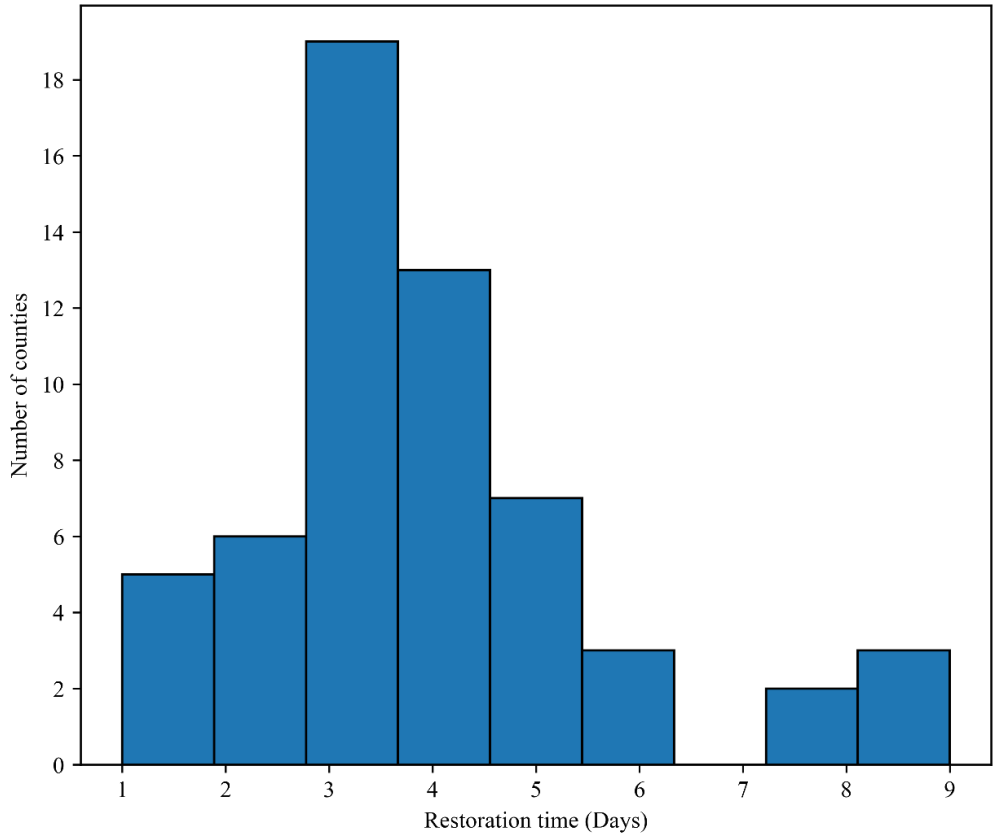

**Fig. 2.** Histogram for restoration times of in Florida after Hurricane Irma

**3.2 Hazard characteristics**

Under hazard characteristics we considered four types of covariates: maximum sustained wind speed, the percentage of power outage in each county, the percentage of census tracts prone to flash flood in each county, and the percentage of census tracts prone to sea level rise in each county. The wind speed for Hurricane Irma was estimated from the HAZUS-MH wind model (Vickery et al., 2000, 2006). This model creates the wind speed profile probabilistically due to a hurricane event. Using this model, we obtained maximum sustained wind speed at census tract level based on their distance to the center of the hurricane. For a given county, the highest wind speed among all census tracts was considered as the maximum sustained wind speed for that county.

We considered the maximum percentage of customers faced power outage from 9$^{th}$ September 2017 to 28$^{th}$ September 2017. We collected this information from Florida Division of Emergency Management (FDEM) as well.

**3.3 Built environment characteristics**

Under built environment characteristics, we considered the percentage of customers served by investor-owned company, and power system variables.

According to Florida Public Service Commission (FPSC), there are three types of electric service providers in Florida: investor-owned electric utilities, rural electric cooperatives, and municipal electric utilities. Investor-owned electric utilities include Florida Power and Light Company, Duke Energy, Tampa Electric Company, Gulf Power Company, and Florida Public Utilities cooperation. Florida also has 34 municipally owned electric utilities and 18 rural electric cooperatives. Investor-owned electric companies are private companies not associated with any government agency. Municipal electric utility companies are public power utilities. These are non-profit utilities owned and operated by state or local governments or agencies. Rural Electric Cooperatives are private, independent, non-profit electric utilities owned by the customers they serve, and tend to provide service in rural areas that are not served by other utilities. In most of the cases, either any two of these three electricity utility companies or all the three types of companies provide services to all the 67 counties in Florida, with varying number of customers across various counties. We considered the percentage of customers served by investor-owned electric utilities in each county. We added this variable for two purposes: i) to understand how these companies responded during the restoration process and ii) if there is any discrepancy in restoration across various electric companies. We collected the number of total customers under each type of electric companies along with total number of customers for each county from FPSC ([psc.state.fl.us/Home/HurricaneReport](psc.state.fl.us/Home/HurricaneReport)), publicly available in their database. However, this data is available from May 2018 and Hurricane Irma occurred in September 2017. So, we estimated how the percentage of customers served by different utilities changes over time. We found that due to population increase the number of customers served by different electric companies did not increase by more than 1%. Moreover, time difference between September 2017 and May 2018 is less than 8 months. Thus, it is reasonable to use the percentage of customers served by the investor-owned companies available for May 2018.

We also consider the number of substations, power plants, and total length of overhead line in each county. They provide a measure of the extent of power system. We collected power system data from U.S. Energy Information Administration, EIA (eia.gov/maps/layer_info-m.php).

### 3.4 Socio-demographic characteristics

As socio-demographic characteristics, we included the median income of the households, and the percentage of non-White American population in each county. We collected the percentage of Hispanic population and median income for the counties from the demographic and economic characteristics of 2013 – 2017 American Community Survey (ACS) 5-Year Data Profile. In Florida 55% of the populations are non-Hispanic, White alone and 27% of the populations are Hispanic. Rest of the populations are Black or African American, Asian-alone and others (*U.S. Census Bureau QuickFacts: Florida*, 2021). To include the poverty information, we used the median household income from ACS.

The descriptive statistics of the data are given in **Table 1**. To understand the presence of correlations among the predictors, Pearson's correlation was calculated (**Fig. 3**). Correlation values among number of power plants, number of substations, and length of overhead power lines are high (0.72 and 0.77) ('Raithel, 2008). So, we considered only the number of power plants among these three variables in the statistical model.

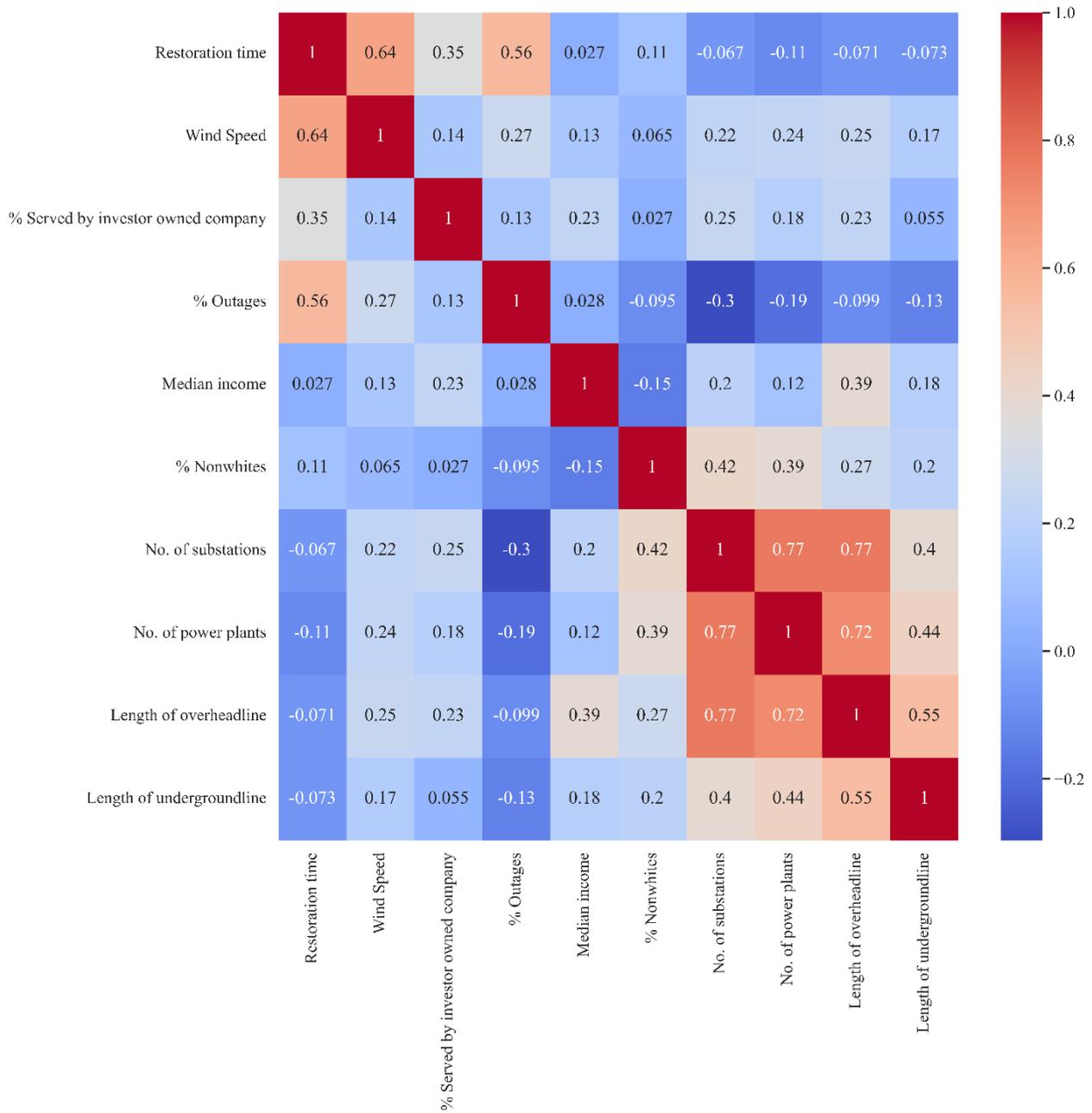

**Fig. 3**. Pearson's correlations between variables

**Table 1.** Data description of continuous variables

|  | Variable | Mean | Std. | Min | Median | Max |
|---|---|---|---|---|---|---|
| **Dependent variable** | Restoration time (Days) | 3.83 | 1.93 | 1 | 3 | 9 |
| **Hazard Characteristics** | Maximum sustained wind speed (mph) | 62.28 | 24.82 | 10 | 64 | 114 |
|  | % of customers who faced power outage | 77.09 | 17.28 | 39 | 78.50 | 100 |
| **Built-environment Characteristics** | % of customers served by investor-owned power company | 55 | 33.9 | 0 | 53 | 100 |
|  | Number of power plants | 3.87 | 4.37 | 0 | 2 | 25 |
|  | Number of substations | 38.82 | 42.15 | 2 | 23.5 | 218 |
|  | Length of overhead line (km) | 1065.62 | 736.96 | 144.52 | 833.43 | 2937.25 |
| **Socio-demographic Characteristics** | % of Hispanic population | 15.03 | 13.22 | 2.4 | 9.9 | 67.5 |
|  | Median Income ($) | 46242 | 9029 | 31816 | 45424 | 73640 |

## 4. METHODOLOGICAL APPROACH

The methodological approach in this study has two main parts. First, we determined the spatial distribution of the restoration time based on the disruption of electricity services. Second, we adopted a statistical modeling approach to determine the factors associated with restoration time from power outages.

### 4.1 Spatial distribution for restoration time of power outage

To identify if there is a clustering pattern between restoration times of electricity disruption in the affected areas, we used Global Moran's I (equation (1)) (Ord & Getis, 1995) which is typically used to estimate spatial autocorrelation. Moran's I was used by Jackson et al., 2021 to understand the spatial trends in county-level COVID-19 cases and fatalities in the United States during the first year of the pandemic.

$$I = \frac{N}{\sum_i \sum_j w_{ij}} \frac{\sum_i \sum_j w_{ij}(X_i - \bar{X})(X_j - \bar{X})}{\sum_i (X_i - \bar{X})^2} \quad (1)$$

where, $w_{ij}$ is the spatial weight having a value of 1 if county $i$ has a shared boundary with another county $j$ or having a value of 0 if otherwise; $X_i$ is the restoration time; and $\bar{X}$ is the average restoration time of all counties considered in the analysis.

Global Moran's I does not tell anything about the places where the patterns are located. The concept of a local indicator of spatial association was suggested to remedy this situation (Anselin, 1995). We applied Local Moran's I (equation (2)) to understand where the clustering patterns are located.

$$I = z_i \sum_j w_{ij} z_j \qquad (2)$$

In equation (2), $z_j$ is the deviation from mean and the summation over $j$ such that only neighboring values are included. Global Moran's I operate by comparing how similar every object (such as a census tract) is to its neighbors, and then averaging out all these comparisons to give us an overall impression about the spatial pattern of variables. On the other hand, Local Moran's I is a local spatial autocorrelation statistic based on decomposition of Global Moran's I statistic. For each observation, it gives an indication of the extent of significant spatial clustering of similar values around that observation. Local Moran's I can reflect total five patterns in spatial clustering (High-High, Low-Low, Low-High, High-Low, and Not Significant). In addition to Local Moran's I, we plotted a choropleth map to visualize the spatial distribution of restoration times. We used ESDA and PySAL packages in Python 3.9 to calculate the Global, and Local Moran's I.

### 4.2 Statistical modeling approach

To determine the effects of the factors (described in Section 3) on restoration times from power outages, we developed a generalized accelerated failure time model (GAFT). We used GAFT model for two reasons: (i) as our dependent variable is restoration time, we considered this to be a time to event data analysis; (ii) as our dependent variable is likely to have a spatial dependency, the GAFT model can be used for modeling both spatial and non-spatial data. To account for the spatial dependence, a random effect (frailty) is introduced into the linear predictor of survival model. Both georeferenced and areally observed spatial locations are handled via random effects (frailties) (Zhou et al., 2020). The GAFT model is given by the following equations (Zhou et al., 2020).

$$S_{x_{ij}}(t) = S_{0,z_{ij}}(e^{-X_{ij}^T \beta - v_i} t) \qquad (3)$$

Or equivalently,

$$y_{ij} = log(t_{ij}) = \tilde{X}_{ij}^T \tilde{\beta} + v_i + \epsilon_{ij} \qquad (4)$$

where $\widetilde{X_{ij}}$ is the matrix of covariates with an intercept term, $X_{ij}^T$ means the transpose matrix of $X_{ij}$, $\tilde{\beta}$ is the vector of corresponding coefficients, $t_{ij}$ is the time, $\epsilon_{ij}$ is a heteroscedastic error term independent of $v_i$, and $S_0(t)$ is the baseline survival function. In the GAFT model, $S_0(t)$ may depend on certain covariates, $z_{ij}$, where $z_{ij}$ is a subset of $X_{ij}$; in this study, we considered $z_{ij} = X_{ij}$. In AFT model, $S_0(t)$ is assumed to be a static parametric survival function, free of covariates. That is, the resulting survival curves are not allowed to vary for different covariates. In practical application, this assumption does not always seem to be true (Hensher & Mannering, 1994). In Generalized AFT model, $S_0(t)$ is allowed to flexibly vary with covariates which has increased the flexibility of the model. Finally, $v_i$ is an unobserved frailty term associated with a county; $i$ indicates the index of an observation (i.e., county) and $j$ indicates the index of a predictor variable.

We estimated this model in R using the spBayesSurv package and the frailtyGAFT function. The detailed description of this package and model can be found in Hsu et al. (2015) and Zhou et al. (2020). As this is a Bayesian modeling approach, it requires to set the prior distributions of the parameters based on domain knowledge. However, this prior knowledge is usually not available (Ulak et al., 2018). In this study, we set most of the prior information according to the default values of frailtyGAFT function under spBayesSurv package in R due to the unavailability of the prior information about the actual parameter distributions and validated it using the trace plots obtained from the model.

The Bayesian specification for prior distribution of the model used in this study is given below (Hsu et al., 2015; Zhou et al., 2020):

$$\tilde{\beta} \sim N(m_0, S_0) \tag{5}$$

$$(v_1, \ldots \ldots v_m)^T | \tau \sim ICAR(\tau^2), \tau^{-2} = \Gamma(a_\tau, b_\tau) \tag{6}$$

$$\epsilon_{ij} | G_{z_{ij}} \widetilde{ind} \, G_{z_{ij}} \tag{7}$$

$$G_{z_{ij}} | \alpha, \sigma^2 \sim LDTFP_L(\alpha, \sigma^2); \alpha \sim \Gamma(a_0, b_0), \sigma^{-2} \sim \Gamma(a_\sigma, b_\sigma) \tag{8}$$

For the coefficients $(\widetilde{\beta})$, a normally distributed prior is considered. For the frailty terms, in GAFT model, a Conditional Auto-Regressive (CAR) prior is chosen for areal data (indicating the spatial data is included based over a geographic area) and a GRF prior is chosen for georeferenced data (indicating the data is included based on coordinates). We chose CAR prior to model the frailty as this study is county-level analysis. Since we included spatial data at a county level, we can assume it as areal referenced data instead of georeferenced data. For areal data, the Intrinsic Conditional Auto-Regressive (ICAR) prior smooths neighboring geographic-unit frailties $v = (v_1, \ldots \ldots \ldots, v_m)^T$. Details on ICAR $(\tau^2)$ prior (equation (6)) is given by the set of conditional distributions in equation (9). Adjacency matrix, E = $[e_{ij}]$ of $m \times m$ dimension for the $m$ regions is used to calculate the frailties, $v_i$. In equation (9), $e_{ij}$ is 1 if counties $i$ and $j$ share a common boundary, 0 otherwise and $e_{ii} = 0$. While calculating $v$ for a region $i$, the other regions under consideration are $j$. $e_{i+} = \sum_{j=1}^{m} e_{ij}$, is the number of neighbors for region $i$ (Zhou et al., 2017, 2020).

$$v_i | \{v_j\}_{j \neq i} \sim N(\frac{e_{ij} v_j}{e_{i+}}, \frac{\tau^2}{e_{i+}}), i = 1, \ldots \ldots, m \tag{9}$$

In GAFT, for spatial analysis, the error term ($\epsilon_{ij}$) is not independent. For this reason, a heteroscedastic error term is introduced over a probability measure $G_z$, defined on $\mathbb{R}$ for every $z \in X$ and a linear dependent tailfree processes (LDTFP) prior is considered for $G_z$. An LDTFP centered at a normal distribution $\phi_\sigma$ is focused with mean 0 and variance $\sigma^2$, that is, $E(G_z) = N(0, \sigma^2)$ for every $z \in X$ (details are described in Jara & Hanson (2011) and Zhou et al. (2017)).

Since the posterior distribution for coefficients of the covariates are unknown, we ran Markov Chain Monte Carlo (MCMC) simulation. For MCMC simulation, we ran 4 chains of 50,000, where 16,000 scans are thinned after a burn-in period of 30,000 based upon examination of trace plots for model parameters (**Fig. 6**). A trace plot is a diagnostic tool for assessing the mixing of a chain. It shows the iteration number against the value of the draw of the parameter at each iteration. It also shows whether chain gets stuck in certain areas of the parameter space, indicating bad mixing.

## 5. RESULTS AND DISCUSSIONS

### 5.1 Result for spatial distribution

We mapped the restoration time (described in section 3.1) with associated county over Florida (**Fig. 4**). **Fig. 4** shows that the southern parts of the Florida (Monroe, Lee, Collier, Charlotte, Broward, Miami-Dade, Palm-Beach, Hendry counties) needed priority during restoration process after Hurricane Irma. In addition, it shows that counties in the middle of Florida (Seminole, Orange, St. Johns, Putnam, Marion) and some in the North (Hamilton, Suwannee, Lafayette) faced moderate (4-7 days) duration of disruption and needs attention for fast recovery.

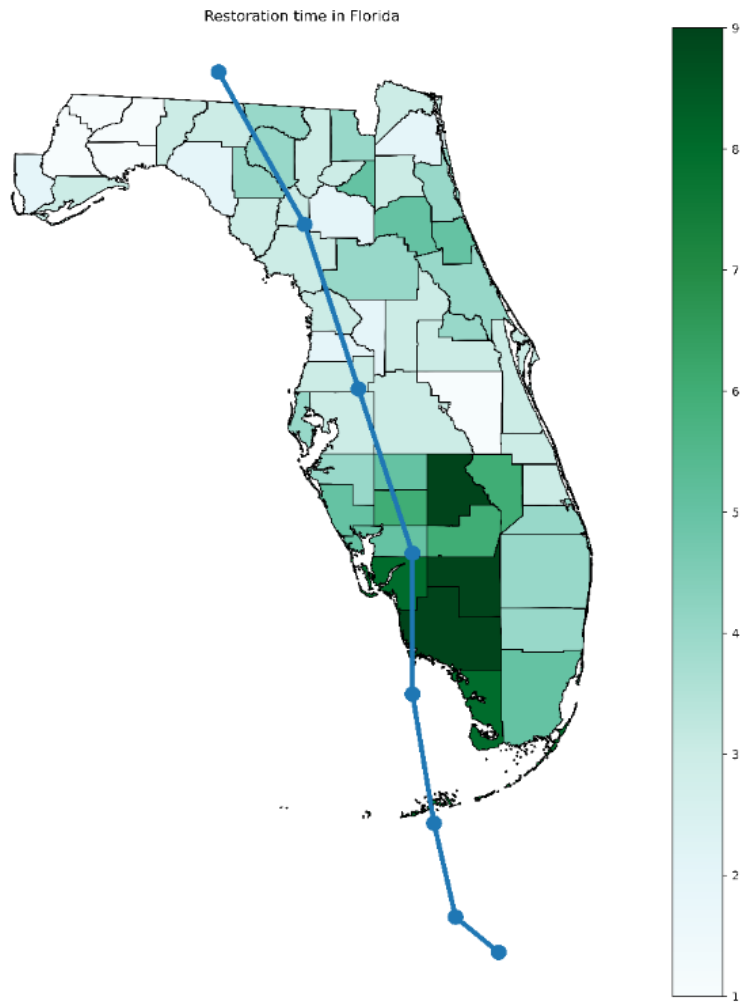

**Fig. 4.** Restoration time of Florida along with the hurricane path (in blue color)

The obtained Global Moran's I value is 0.58 (*p*-value = 0.001) indicating the presence of spatial autocorrelation. **Fig. 5** shows locations of the clustering patterns for the restoration time from power outages. The Global Moran's I test within the entire study area shows significant ($p < 0.05$) spatial autocorrelation for our target attribute.

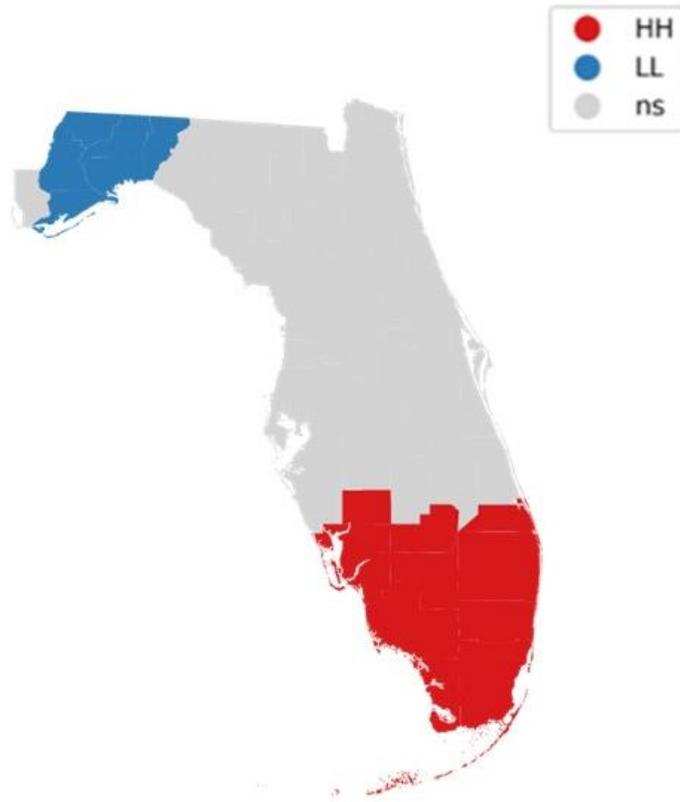

**Fig. 5.** Local Moran's I plot for restoration time of power outage.

Local Moran's I plot (**Fig. 5**) shows the clusters of longer restoration time (hot spots) and the clusters of short restoration time. The local Moran's I test shows considerable spatial clustering for 17 counties (local clusters are significant, p < 0.05, **Fig. 5**). The grey areas in **Fig. 5** are the locations where no significant spatial patterns were found; the red areas are the counties where people had longer restoration time living closely to other counties with longer restoration time. The low with low (L-L) are all the blue areas, those are locations where people had shorter restoration time living closely to other counties with shorter time of restoration process. For Hurricane Irma in Florida, we could not find any HL or LH clustering pattern.

## 5.2 Result from statistical analysis

For statistical analysis, we considered spatial models because of the obtained Global Moran's I statistics found in section 5.1. A high Moran's I value of 0.58 (*p*-value: 0.001) clearly indicates the presence of spatial correlations among observations. As such, a non-spatial model assuming independent and identically distributed (i.e., IID) observations, ignoring spatial correlations, will not be appropriate. Spatial survival analysis is used to analyze clustered time to event data when the clustering issue arises from geographical regions (Banerjee, 2016).

Table 2 presents the results of the Generalized Accelerated Failure Time (GAFT) model. Two separate models were fitted with and without considering spatial correlation. The proposed GAFT model with CAR frailties has the larger log-pseudo marginal likelihood (LPML) (-74) compared to the non-frailty GAFT model (-83), indicating that considering spatial correlation improves the model fit by 12%.

Bayes factors is a Bayesian alternative to classical hypothesis testing. The Bayes factors for testing all the covariates' effects on baseline survival were found to be greater than 100, indicating that the baseline survival function (equation (3)) under the AFT model depends on these variables, and thus GAFT model should be considered (Zhou et al., 2020). The mean posterior inference of conditional CAR frailty variable was found to be 0.212, representing the amount of spatial variation across counties. The trace plots of the regression coefficients (Fig. 6) have even and stationary pattern, indicating that the MCMC simulations converged (Zhou et al., 2020).

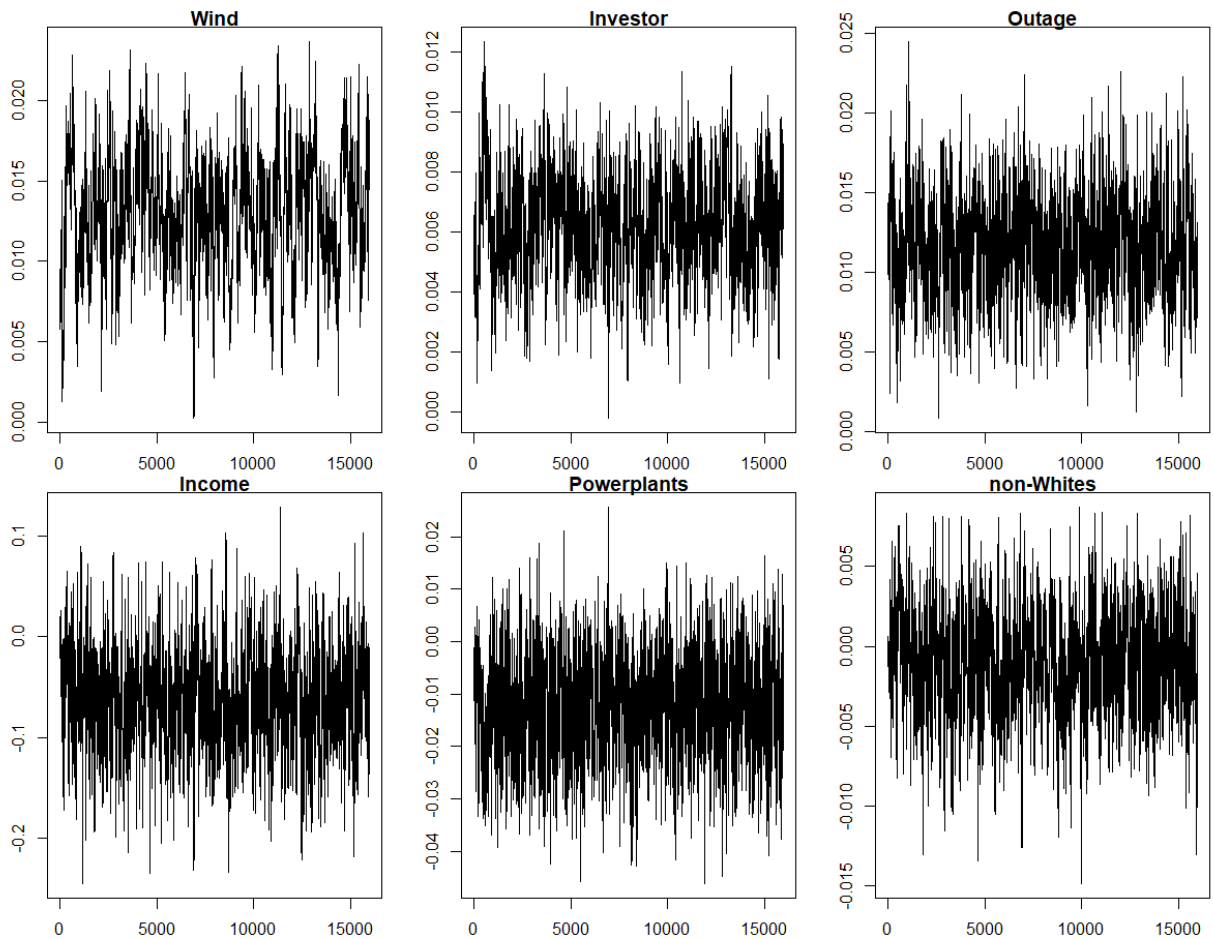

**Fig. 6.** Trace plots of regression coefficients

Standard deviations of the maximum sustained wind speed, the percentage of customers served by investor-owned power companies, the percentage of customers faced power outages, and the number of power plants are small compared to the mean (Table 2). Moreover, 90% high posterior density interval of the regression coefficients do not contain zero, indicating that these four variables have significant influence on restoration time.

**Table 2.** Posterior inference of regression coefficients

|  | Model with CAR frailties | | Model without CAR frailties | |
|---|---|---|---|---|
| **Variable** | **Mean (Std. dev)** | **90% HPD** | **Mean (Std. dev)** | **90% HPD** |
| **Intercept** | -0.746 (0.347) ** | [-1.323, -0.192] | -0.759 (0.224) ** | [-1.121, -0.389] |
| **Maximum sustained wind speed** | 0.013 (0.003) ** | [0.007, 0.019] | 0.011 (0.0016) ** | [0.009, 0.014] |
| **% of customers faced power outage** | 0.0117 (0.0029) ** | [0.007, 0.017] | 0.012 (0.0029) ** | [0.007, 0.017] |
| **% of customers served by investor owned company** | 0.0062 (0.002) ** | [0.003, 0.009] | 0.006 (0.0012) ** | [0.004, 0.007] |
| **Number of power plants** | -0.015 (0.007) * | [-0.03, -0.0007] | -0.023 (0.009) ** | [-0.041, -0.008] |
| **Median Income** | -0.072 (0.040) * | [-0.13, -0.00016] | -0.062 (0.042) * | [-0.12, -0.0009] |
| **% of Non-White population** | -0.001 (0.003) | [-0.006, 0.004] | 0.003 (0.002) | [-0.001, 0.007] |
| **Log pseudo marginal likelihood** | -74 | | -83 | |
| ***Significant at the 90% highest posterior density interval.** | | | | |
| **** Significant at the 95% highest posterior density interval.** | | | | |

Among hazard characteristics, the maximum sustained wind speed and the percentage of customers faced power outages were found to be significant and positively associated with power service restoration time. A positive association means that an increase in a predictor variable will increase restoration time and a negative association indicates the opposite. The exponentiated coefficient of maximum sustained wind speed ($e^{0.013} = 1.013$) is the factor by which the mean restoration time increases by 1.3% with one mph increase in maximum sustained wind speed. One percent increase in % of customers without power ($e^{0.012} = 1.012$) increases the mean restoration time by 1.2%. Percentage of census tracts prone to flash flooding had positive association with restoration time too, but this variable was not found to be significant in the spatial model.

Among built-environment characteristics, percentage of customers served by investor-owned power companies was found to be significant and positively associated with power service restoration time. In addition, the number of power plants was also found to be significant. Among socio-demographic variables, median income was found to be statistically significant.

## 6. DISCUSSION

In this paper, we studied how hazard, built environment, and socio-economic characteristics of a region are associated with restoration time of power outages due to a hurricane. Our results indicate that counties with higher wind speed had longer restoration times. It is likely that high wind speed during Hurricane Irma caused greater damages to the electric infrastructure systems causing a longer restoration time. The positive coefficient for the percentage of customers faced power outage indicates that for regions where higher percentage of customers were out of electricity, it took longer time for the maintenance teams to restore power service in such places.

The percentage of customers of a county served by an investor-owned utility company is also positively associated with restoration time. It indicates that counties with a higher percentage of customers served by investor-owned electric companies, faced longer restoration time, adjusting for other covariates and county of residence. This may have happened because the regions where most of the households are served by investor-owned utility companies, also faced higher wind speed, and had a large number of customers with power outages. As a result, it took long time for the investor-owned power companies to restore electricity disruption.

The number of power plants is negatively associated with restoration time, adjusting for other covariates and county of residence. That is, counties with more power plants, were able to restore their power services fast. A greater number of power plants indicates a more extensive and better power system of a region. In other words, these areas are prioritized to get more systems up and running, resulting in a shorter restoration time of power outages. Utility companies might have prioritized restoration in regions with large number of power plants since component-based restoration strategies prioritize critical components in the following order: power plants, substations, transmissions, and distributions (Esmalian et al., 2022). Moreover, we found that it took a longer time for investor-owned power companies to restore electricity disruption, perhaps, because of a high number of outages present in the regions served by investor-owned companies. Hence, instead of a component-based restoration strategy, an outage-based restoration strategy can be prioritized focusing the regions with a greater number of customers without power. Population and vulnerability based restoration strategies were found to be better than a component based strategy in the agent based simulation by Esmalian et al., (2022).

Fig. 7 highlights counties with significant factors of longer power service restoration time using county level data. For example, maximum sustained wind speeds in South-West counties of Florida (Monroe, Collier, Lee, Hendry, Highlands) were greater than South-East counties (Miami-Dade, Broward, Palm Beach, Martin, St. Lucie) and North-West counties (Taylor, Jefferson, Leon, Wakulla, Gadsden, Gadsden, Liberty, Franklin). As a result, South-west counties on average (8 days) had longer time of power outage, South-East counties faced on average 4.5 days, and North-West counties on average 1.75 days of power disruption. Counties where 75% or more customers were served by investor-owned power companies on average faced 4.75 days of electricity disruption. Collier and Highlands counties faced 9 days of power disruption where about 87% of the customers were served by investor-owned power companies. In such counties, the mean percentage of customers who faced power disruption was also higher (79%). In Collier and Highlands, about 97% customers lost power services due to Hurricane Irma. Counties with 4 or more number of power plants (Polk, Leon, Hillsborough, Alachua, Orange, Osceola) on average faced 3.5 days of power disruption.

Previous studies on Hurricanes Irma (Mitsova et al., 2018) and Hurricanes Bonnie, Isabell, Dennis, and Floyd (Liu et al., 2007) show that maximum sustained wind speed is positively associated with power service restoration time. Number of power plants is important to predict thunderstorm induced power outages (Kabir et al., 2019). Mitsova et al. (2018) found longer disruption for municipal owned power companies and rural cooperatives. Besides, they found the percentage of Hispanic population to be significant, which contradict with our results. One possible reason for these discrepancies could be that Mitsova et al. (2018) considered wind speed information as a dichotomous variable, which cannot account for the differences of wind speeds across counties. Thus, the effect of wind speed on restoration times are captured by other variables (e.g., % of customers served by different power companies, % Hispanic population, etc.). On the contrary, we have considered actual maximum sustained wind speed for each county. It is often assumed that poor, minority communities are less prioritized reflecting inequality in power service restoration activities. Previous studies also found disparities in experienced hardship due to power outages in Puerto Rico and Texas during Hurricane Maria and Harvey (Azad & Ghandehari, 2021; Coleman et al., 2020). Consistent with these studies, we found disparity issue with respect to median income for power restoration time in Florida during Hurricane Irma. This necessitates accelerated recovery activities and better infrastructure systems in low-income communities to make them resilient to hurricane impacts.

Based on the significant factors (e.g., maximum sustained wind speed, % of customers faced power outage, % of customers served by investor-owned power companies, and number of power plants) obtained from the GAFT model, areas likely to face a longer disruption time after a hurricane can be identified. For most of the counties, these four variables could capture the possible critical regions for restoration process of power outages (**Fig. 7**).

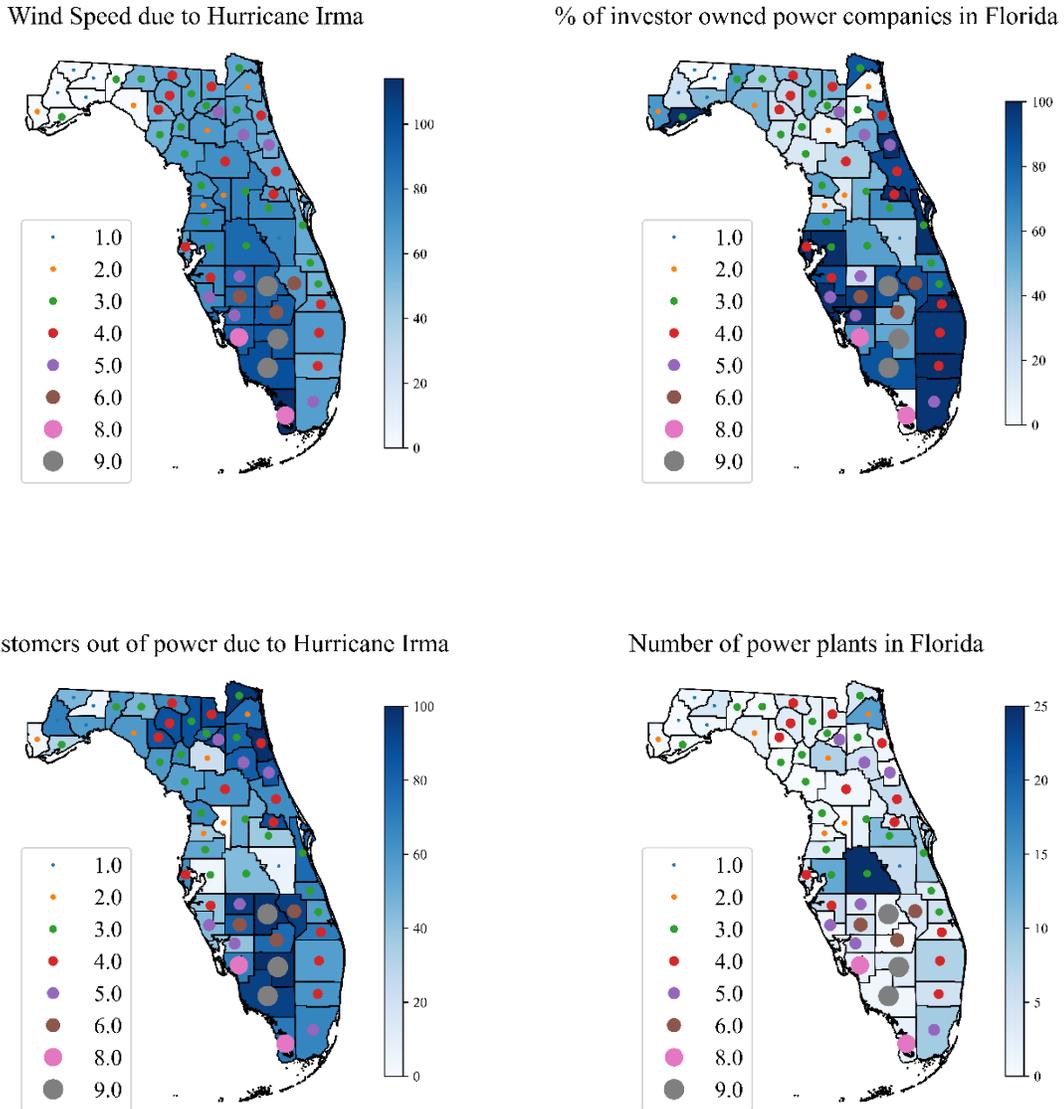

**Fig. 7.** Counties in Florida mapped by significant variables for power service restoration time (color bar represents the factors and dots represents the restoration time in days)

## 6. CONCLUSIONS

In this study, spatial distribution of restoration time was investigated at a county-level to identify less resilient location for electricity disruption. We present a Generalized Accelerated Failure Time (GAFT) model to determine the factors which have impacts on electricity infrastructure systems. Considering spatial correlation in time to event data analysis has improved the model fit by 12% compared to the model without considering spatial correlation. The proposed model holds potential for the analysis of power service restoration time due to extreme events as it can consider spatial clustering particularly for time as a dependent variable. We considered three types of factors: hazard characteristics, built environment characteristics, and socio-demographic characteristics in the proposed GAFT model. This model showed that total five factors among six were associated with restoration time of power services caused by Hurricane Irma. We have found that factors such as sustained wind speed, percentage of customers facing power outage, percentage of customers served by investor-owned power company, median household income and number of power plants are strongly associated with restoration time. The findings of this study suggest that counties with a higher percentage of customers served by investor-owned electric companies, few number of power plants and lower median household income, faced power outage for a longer time. Hence, recovery strategies based on number of outages and vulnerability (in terms of median income) may improve the overall power outage recovery time.

The described approaches and finding of the study can aid policy makers and emergency officials in understanding factors that should be given importance during the restoration process after a hurricane. This study will also allow them to identify which critical counties or regions need attention for restoration process and can ensure rapid restoration and minimize losses in the affected regions. In general, electricity companies have the knowledge about power system variables (e.g., number of power plants, substations, and total length of overhead) and number of outages but do not have much knowledge about disaster conditions. Therefore, if utility companies can work with emergency managers to understand the relationship between disaster condition and electricity disruption, they could take necessary steps that would account for disaster conditions. Such efforts can improve electrical grid resilience during extreme events and lead to improved recovery outcomes. Pre-disaster preparation and risk mitigation programs necessitate better electricity facilities. To ensure a faster restoration of power systems due to an extreme event, an increased awareness of both hazard and built environment characteristics is required by utility companies, emergency management professionals, and policy makers. This study highlights the need for considering spatial dependence among observations to better understand how power restoration after hurricanes is impacted by hazard and built environment characteristics of the affected regions.

Most previous studies (Kabir et al., 2019; Liu et al., 2007) were based on proprietary data from utility companies. This does not allow reproducibility of the research and prevents implementation in actual crisis management. All the factors included in this study were collected from publicly available data. For example, projected hurricane path or wind speed information can be obtained from National Weather Service (NWS) and National Hurricane Center (NHC) when planning for power restoration before a hurricane strikes. Similarly, socio-economic characteristics of a community are available in ACS. Thus, the variables used in this study can be easily collected and utilized before the occurrence of a hurricane to predict restoration time. Such predictions will help policy makers and emergency officials to accelerate the overall restoration process from power outages.

However, our analysis has several limitations which include: this study is a county-level analysis for power service restoration time. However, county is not the finest geographic unit. In future, focus can be given at smaller level of geographic units (e.g., county subdivision, zip code or census tracts) based on data availability. It is important to note that these limitations are related to data availability and not to the proposed methodological approach. These limitations can be overcome if relevant agencies such as utility companies share outage data and hurricane wind speed projections are available at a higher resolution.

## DATA AVAILABILITY STATEMENTS

All the data are available online. Models, or code that support the findings of this study are available from the corresponding author upon reasonable request.

## DECLARATIONS

**Conflict of interest:** The authors declare no competing interests.

**Ethical statement:** All procedures were carried out in conformity with the necessary rules and guidelines. No Institutional Review Board protocol was required for this analysis because it was an observational study of aggregate-level data.